%% file: 00_starthere.tex
\documentclass[11pt]{article} 

\usepackage{times} 
\usepackage{bm} 
\usepackage{color} 
\usepackage{hyperref} 
\usepackage{graphicx} 
\usepackage{vmargin} 
\usepackage{afterpage}
\usepackage{pdfpages}
\usepackage{xurl}
\usepackage{url}
\definecolor{green}{RGB}{11,155,13}

\hypersetup{colorlinks=true,linkcolor=blue,urlcolor=blue}
\setpapersize{USletter}
\setmarginsrb{1in}{1in}{1in}{1in}{0pt}{0mm}{0pt}{0mm}

\newcommand{\pseudodot}{{\lower 2.4pt\hbox{$\cdot$}}}
\usepackage[utf8]{inputenc}
\usepackage{tabularx}	
\usepackage{multirow}	
\usepackage{booktabs}

\begin{document}

\setlength{\baselineskip}{14pt} 
\setlength{\normalbaselineskip}{12pt} 

\section*{Understanding ADHD Productivity in Construction Work: \\Toward AI-enabled VR Interventions}

\vspace{0.7em}

ZINAT ARA, \small{\textit{Information Sciences and Technology, George Mason University, zara@gmu.edu}} \\
BEHZAD ESMAEILI, \small{\textit{School of Industrial Engineering, Purdue University, besmaei@purdue.edu}} \\
LAP-FAI YU, \small{\textit{Computer Science, George Mason University, craigyu@gmu.edu}}\\
SUNGSOO RAY HONG, \small{\textit{Information Sciences and Technology, George Mason University, shong31@gmu.edu}}

\vspace{1em}

\paragraph{Abstract}
Attention-Deficit/Hyperactivity Disorder (ADHD) is identified as the most prevalent neurodivergent condition in the construction industry. While the construction industry may broaden employment opportunities, little is known about how ADHD traits shape workers’ performance, sustained attention, and situational awareness in dynamic job-site environments. This work presents an exploratory interview study aimed at understanding how ADHD traits influence construction-specific productivity and how future interventions can reduce challenges while amplifying strengths. We conducted semi-structured interviews with construction workers with ADHD, safety managers, and ADHD researchers to capture their perspectives on attentional demands, task coordination, and workplace adaptation. As part of these discussions, participants also reflected on the potential of combining artificial intelligence (AI) and virtual reality (VR) to support future ADHD workers. Our analysis identifies two overarching themes: (1) workplace challenges, capturing difficulties that arise from both the nature of construction work and the specific needs of ADHD workers, and (2) productivity support strategies, describing effective methods to sustain focus and task engagement. Further, we derive design requirements for AI and VR-enabled interventions that provide adaptive attentional scaffolding, mediated social presence, and motivational support. We conclude by discussing how these insights can inform the future development of ADHD productivity enhancement support. 

%
%

\input{sections/01_introduction}

\input{sections/02_relatedwork}

\input{sections/03_method}
\input{sections/04_result}

\input{sections/05_design}

\input{sections/06_conclusion}

\vspace{0.4em}
\paragraph{Acknowledgments} The authors gratefully acknowledge support from the National Science Foundation (NSF \#2310210 and \#2418236) for the study. 

\bibliographystyle{IEEEtran}
\bibliography{sections/99_new_ref}

\end{document}

%% file: sections/01_introduction.tex
\section{Introduction}


Attention-Deficit/Hyperactivity Disorder (ADHD) is reported as the most prevalent neurodivergent condition in the construction industry, representing approximately 54\% of neurodivergent construction workers~\cite{nfb2023neurodiversity, pbctoday2023neurodiversity, ara2026rethinking}.
Many individuals with ADHD are drawn toward the field due to its alignment with their cognitive styles~\cite{nfb2023neurodiversity,adhdcons}.
Unlike low-stimulation desk jobs, construction environments are dynamic, physically engaging with hands-on concrete tasks that can reduce boredom-driven distraction~\cite{doyle2020neurodiversity,barkley2015executive} for them.
At the same time, working in the construction industry requires frequent attentional shifts and continuous situational awareness, which are critical for both productivity and safety~\cite{hasanzadeh2017impact,chang2025impacts}.
However, workers with ADHD often experience a distinct pattern of attention, where focus is highly context-sensitive.
These traits can influence how they engage with tasks, including navigating multiple demands, prioritizing actions, managing time, and regulating emotional responses~\cite{campbell2023adhd,barkley2015executive, ara2023exploring, feng2026lost}.
These characteristics highlight both the strengths and challenges of ADHD in construction contexts, pointing to the need for systems that can dynamically align with workers’ attentional patterns rather than impose rigid structures.
In response to this, Artificial Intelligence (AI) offers promising outcomes to support neurodivergent workers through adaptive and personalized assistance in the workplace~\cite{holmes2019artificial, schuller2021ai,ara2024collaborative,wheeler2018navigating, zhou2026empowering}.
Similarly, immersive technologies such as Virtual Reality (VR) have been widely adopted in construction engineering education, training, and worker safety programs to address critical safety challenges~\cite{yu2026col,wang2018critical,barnard2024neurodiverse}. 
Artificial Intelligence (AI) has wide applicability across diverse domains~\cite{ara2024closing, ara2022predicting, ara2021traffic, khan2024aurora, mim2025investigation, mim2025exploratory} 
Therefore, building on these advancements, integrating VR with AI presents strong potential to support sustained attention and enhance performance for workers with ADHD. 
AI-enabled VR systems can further strengthen workplace support by providing interactive environments, real-time feedback, and context-aware interventions within dynamic and safety-critical construction settings.

Despite the potential, there remains a significant gap in understanding the \textit{``lived experiences''} of construction workers with ADHD and how such technologies can effectively support them.
Prior research in construction has primarily explored VR for safety training and skill development~\cite{jonescu2024understanding}, teamwork effect~\cite{wang2018critical,hasanzadeh2017impact}, hazard recognition~\cite{chang2025impacts}, and learning outcomes~\cite{li2018critical} in controlled environments.
A recent study shows that workers with ADHD face difficulties with attention regulation, task management, and safety compliance, highlighting the need for greater awareness and workplace accommodations in the mining industry~\cite{ziolkowska2025challenges}.
However, these insights are not directly transferable to construction contexts, where work is highly dynamic, collaborative, and safety-critical. 
This underscores the need for more domain-specific investigation.
Existing research provides a limited understanding of the challenges and coping strategies of workers with ADHD, as well as how AI and VR systems can support their attention and productivity in construction, an area that remains underexplored within HCI and CSCW literature.

To address the gap, this work investigates the challenges faced by workers with ADHD in construction settings, the strategies that can support their workplace performance, and the opportunities for designing AI-enabled VR systems to enhance attention and productivity.
We conducted a qualitative interview study with six participants from diverse stakeholder groups, including construction workers with ADHD, safety managers, and ADHD researchers.
Our overarching Research Questions (RQs) are as follows:

\begin{itemize}
    \item \label{rq1} \textbf{RQ1.} What challenges do workers with ADHD face in maintaining sustained attention in construction work, and what strategies help them stay engaged and focused on tasks?

    \item \label{rq2} \textbf{RQ2.} How can VR and AI be integrated to provide context-aware, adaptive support to improve attention and overall performance in construction environments?
\end{itemize}

%% file: sections/02_relatedwork.tex
\section{Related Work}

Research on ADHD across health, HCI, and social sciences has identified key factors affecting workplace performance. 
Studies show that even with existing tools, individuals with ADHD continue to face challenges such as task prioritization, time estimation, and task switching~\cite{campbell2023adhd}, largely due to impairments in executive functioning~\cite{nadeau2005career}. 
Both internal factors (e.g., motivation and focus control) and external factors (e.g., work environment and social support) play an important role in shaping their productivity and success at work~\cite{clements2006creating, nadeau2005career}.
Another line of research has explored technological interventions to address ADHD-related challenges, including smartphone reminders, wearables, EEG devices, and AR/VR systems targeting attention and task management~\cite{spiel2022adhd,rajarajeswari2024attention,cuber2024examining}.
Much of this work focuses on attention assessment and improvement in controlled settings, such as reading tasks or virtual environments~\cite{asiry2018extending,rajarajeswari2024attention,cuber2024examining}.
For instance, a study was designed to assess how variations in text color (i.e., highlighting, contrast, and sharpening) affect the reading attention span of children with ADHD~\cite{asiry2018extending}. 
Another study evaluated the effectiveness of a VR-based attention assessment tool that measured attention profiles by asking users to locate a target shape amid various distractions~\cite{rajarajeswari2024attention}. 
Cuber et al. investigated the potential of VR to increase attention span in a study with 27 university students with ADHD, who were asked to complete homework in a quiet virtual reading environment~\cite{cuber2024examining}.
While these approaches offer useful support, they are largely centered on individual coping and do not reflect the complex, collaborative, and safety-critical demands of real-world environments.

Construction sites are fundamentally different from the controlled environments studied in prior ADHD research.
Due to its complex nature, workers require continuous attention shifting, hazard monitoring, and coordination with team members~\cite{hasanzadeh2017impact}.
Few studies have examined how to measure attention in construction contexts, for example, Hasanzadeh et al. empirically assessed workers’ attention using eye-tracking technology~\cite{hasanzadeh2017measuring}.
Other studies have focused on supporting safety in hazardous situations, showing that intense focus on a primary task can cause workers to overlook critical hazards, reducing situational awareness and increasing safety risks~\cite{lee2022assessing}.
For instance, Chang et al. developed a virtual bricklaying game and indicated that ADHD workers exhibited a lower situational awareness of dynamic objects by focusing on the primary task, with no increase in productivity~\cite{chang2025impacts}.
In addition, construction work is highly collaborative, where peers influence risk perception and decision making~\cite{ten2016and}.
This can further complicate attentional demands for ADHD workers in ways that are not captured in existing literature.

VR/AR has received a considerable amount of attention within the research and construction industry in the past two decades~\cite{li2018critical}.
In construction contexts, VR has primarily been designed for safety training and skill development in controlled settings~\cite{jonescu2024understanding, MVST}, demonstrating improvements in hazard recognition and learning outcomes.
Wang et al. conducted a survey study reviewing VR/AR applications in construction by analyzing journal publications from 1997 to 2017~\cite{wang2018critical}. 
Their findings indicate that most VR/AR applications have focused on developing frameworks for construction safety management and visualization systems.
Recently, one study highlights attention regulation challenges and the required psychological and organizational support for individuals with ADHD in the mining industry~\cite{ziolkowska2025challenges}. 
However, these studies do not examine how such challenges unfold during real task execution in dynamic, team-based, and risky construction environments, or how workers actively cope with these demands in practice.
This reveals a critical gap: very little research has investigated the \textit{``lived challenges''} of construction workers with ADHD, the strategies they use to cope in the field, and the factors that positively influence their productivity and safety.
As both AI and VR offer strong potential to address workplace challenges, further research from an HCI and CSCW perspective is needed to understand how AI-enabled VR systems can provide context-aware, adaptive support tailored to these environments.

%% file: sections/03_method.tex
\section{Method}
To capture multi-faceted perspectives on ADHD-related workplace challenges in construction, we conducted open-ended semi-structured interviews with six participants from different expertise.
These participants were well-positioned to inform us about both (1) the nature of construction work characterized by hazardous, high-risk environments and (2) the challenges faced by ADHD individuals, such as difficulties in sustaining attention and situational awareness.
We also sought their perspectives on the opportunities and limitations of VR combined with AI as a medium for productivity support for ADHD workers.
\subsection{Participants Recruitment}
We recruited two neurotypical (NT) construction safety managers with experience supervising workers with ADHD, two construction workers with clinically diagnosed ADHD, two ADHD behavioral expert researchers, one of whom was also diagnosed with ADHD.
As ADHD is a spectrum condition, individuals may present with hyperactivity, inattentiveness, or a combination of both. 
Accordingly, our study recruited high-functioning adults with ADHD who could communicate effectively and perform tasks independently.
Participants were recruited through multiple channels, one of the authors leveraged prior connections with construction personnel to identify potential participants. 
We also shared a recruitment post in online groups for ADHD construction workers and used snowball sampling through referrals. 
Interested individuals received an email with the study procedures, including details about the study purpose, eligibility criteria, compensation (a \$25 gift card), and contact information.
Among the six participants, five had prior VR experience (M1, M2, and R1 used VR in work or research contexts, while W2 and R2 had used it for personal purposes).
Participants ranged in age from 24 to 65 years; two identified as females, and the remainder as males.
Both safety managers held college degrees. 
W1 had a specialized field diploma, while W2 was a high school graduate. 
Both researchers held PhD degrees.
Table 1 summarizes the details of our participants.

\renewcommand{\arraystretch}{1.2}%
\begin{table*}
\small
\centering
 \begin{tabularx}{\textwidth}{p{0.03\textwidth} p{0.26\textwidth}  p{0.04\textwidth} X}
\toprule
\textbf{PID} & \textbf{Types} & \textbf{Age} & \textbf{Profile (years of experience)} \\
\midrule
M1 & Safety Manager (NT) & 52 & Safety Director, managing risk from safety standpoints (29 years) \\
M2 & Safety Manager (NT) & 53 & A Senior Safety Manager who develops safety protocols (25 years) \\
W1 & Construction worker (ADHD) & 65 & An electrician tech in building maintenance (42 years) \\
W2 & Construction worker (ADHD) & 24 & A helper in construction plumbing service (2 years) \\
R1 & ADHD Researcher (NT) & 41 & Researcher and Licensed Psychologist in ADHD (10 years) \\
R2 & ADHD Researcher (ADHD) & 28 & Researcher and Clinical Psychologist (4 years) \\
\bottomrule
\end{tabularx}

\vspace{2mm}

\caption{
A list of participants, from the left: (1) Participant ID, (2) Stakeholder type, (3) Age, and (4) their profile and years of experience.
}

\label{tab:table1}
\end{table*}

\subsection{Interview and Analysis}
All interviews were conducted via Zoom. 
After obtaining informed consent, we recorded the sessions and used a role-specific slide deck to ensure consistency.
We structured the interview questions for safety managers and construction workers with ADHD into the following topics: 
(1) general work processes and safety concerns on construction sites,
(2) ADHD workers' challenges and coping strategies,
(3) practices of collaboration and teamwork, 
(4) key factors influencing workers’ performance in the field. 
In addition, we presented two short video clips from open-source VR construction sites~\cite{Pixo, MVST} that simulated field tasks and safety training scenarios to elicit (5)  their reflection about VR in this context and desired features to support workers’ performance in real-world settings.
For the ADHD researchers, the interview questions centered on the following topics: (1) challenges that workers with ADHD encounter in the work context,
(2) coping strategies workers may find helpful
(3) expert point-of-view of approaches to help ADHD individuals sustain their attention and task performance, and
(4) technology and potential designs that could better assist the workers.
The interviews lasted 50.7 minutes on average (SD=9.01) while the longest one lasted 62 minutes and the shortest one 42 minutes.
All participants were based in the United States and identified as native English speakers. 
Interviews were transcribed by English-proficient transcribers without the use of AI-based transcription tools.
After data collection, we applied an iterative coding process following a reflexive thematic analysis approach~\cite{braun2021thematic}. 
Two coders independently coded text segments from each transcript and met regularly to compare interpretations, resolve discrepancies, and iteratively refine the coding scheme.
Although our sample size was relatively small (n = 6), it intentionally included a diverse set of stakeholders (e.g., construction workers, safety managers, and ADHD researchers) to capture a range of perspectives on ADHD and productivity in construction contexts. 
We conducted data collection and analysis concurrently and monitored for thematic saturation, defined as the point at which additional interviews did not yield substantially new themes~\cite{guest2006data}. 
While some variation in perspectives was expected given the heterogeneous sample, subsequent interviews primarily reinforced previously identified patterns, offering limited novel insights. 
Therefore, our analysis reached thematic sufficiency for identifying recurring cross-cutting themes, consistent with qualitative research practices for exploratory, multi-stakeholder studies~\cite{braun2021thematic}.
The coding process led to a unified thematic structure that was reflected in the Results section.

\subsection{Positionality}
We acknowledge the positionality of the research team. 
None of the authors identify as ADHD, though several have prior experience working with neurodivergent populations in educational, coaching, and research contexts. 
We recognize that this positionality shapes study design, data interpretation, and theme development. 
To mitigate bias, we drew on prior ADHD research, insights from formative studies with ADHD workers, and iterative feedback from neurodivergent advisors. 
All studies were conducted in the United States with adult participants residing there during the study period.



%% file: sections/04_result.tex
\section{Results}

We present findings from three stakeholder perspectives based on our RQs.
Our analysis identified two overarching themes:
(1) \textbf{Workplace Challenges}, capturing difficulties that arise from both the nature of construction work and the specific needs of ADHD workers;
(2) \textbf{Productivity Support Strategies}, describing effective approaches for sustaining attention and enhancing productivity among workers with ADHD, including both real-world strategies and desired relevant VR-based features. 

\subsection{\textbf{Workplace Challenges}}

\subsubsection{\textit{Decision Making under Pressure}}Workers with ADHD often face challenges in construction settings where unpredictability is common. 
While tasks are typically pre-planned, participants emphasized that unforeseen scenarios regularly arise that demand quick decisions. 
As M1 noted, ``\textit{It becomes difficult when being spontaneous occurs for folks in the field. We've pre-planned well, and what our folks wind up facing out there is not a typical scenario}''.
ADHD workers may find it harder to balance the choice between pausing work to make the ``right'' decision or rushing to complete a task under pressure.
M2 described this dilemma: ``\textit{If we haven't prepared our folks to pre-plan then, when they are faced with either stopping work, to do the right thing in the moment, or hurry up and get something done. That's usually the crux of that decision-making process}''.
R2 explained that even when instructions are clear, workers must move beyond compliance into active judgment, which is challenging.

\subsubsection{\textit{Attention and Cognitive Load}} Participants described how difficulty maintaining focus led to misunderstandings, particularly when they had to follow multiple instructions simultaneously. 
This challenge was especially critical in high-stakes tasks requiring precision, where even small lapses in attention could result in errors.
For instance, while working under his supervisor, who is a mechanic at a construction site, W2 mentioned, ``\textit{The biggest problem I remember was that my lead mechanic would give me a measurement, and after a while I would forget and have to ask again, if I'm cutting 2-inch or 4-inch pipe? And he would be like, 'what the hell are you cutting over here? you pay attention stupid!'}''.
R1 remarked that distraction and inattentiveness slow task completion and increase error rates, making it challenging to maintain a balance between the two.
When attention is split across multiple steps, that makes it difficult to maintain working memory of what comes next.
As W1 said, \textit{``If I'm doing a bunch of things at the same time, I kind of stack these things on top of each other, and it's pretty hit or miss, like, sometimes I'll forget the second step and skip to the third step}''. 
Forgetting a step or skipping ahead is not simply a matter of carelessness but a direct result of an overloaded cognitive system struggling to juggle simultaneous demands.

\subsubsection{\textit{Procedural Adaptivity}} For individuals who rely on routine and predictability, sudden procedural shifts disrupt established work patterns and create uncertainty about how to proceed efficiently.
As M2 said, ``\textit{Increased safety protocol has become a multiple-step process, that can be a little bit unnerving for people who did not expect changes}''.
Tasks that push workers outside their comfort zones cause stress, as W2 referred, ``\textit{When there’s a situation where I don’t feel safe doing it, I’m just not doing that until somebody else does}''.
W1 and R2 also mentioned that frequent and strict deadlines intensify the pressures of construction work.

\subsubsection{\textit{Safety Culture Stagnation}} In the construction industry, this translates into potential safety lapses, such as overlooking moving equipment or failing to adjust to unexpected changes in the worksite.
When serious near-miss incidents fail to prompt a paradigm shift or meaningful procedural changes, these incidents raise significant concerns about workplace safety (W2).
M1 emphasized the awareness, ``\textit{So the extreme nature of the moving parts of a project all the time, it's about awareness, which is a key factor in ensuring someone's safety}''.
Workers with ADHD are also more likely to experience “tunnel vision” or hyperfocus, which can reduce their capacity to notice surrounding risks~\cite{fleming2012developmental}.
While this state may temporarily improve productivity, people can lose track of time and fail to notice critical environmental cues.
As R1 mentioned,  ``\textit{Core symptoms of ADHD include difficulties controlling attention, such as being easily distracted or getting hyperfocused. Hyperfocus can prevent multitasking or recognizing environmental cues for needs to shift}''.

\subsubsection{\textit{Coordination Gaps}} Clear communication in construction goes beyond just giving instructions, it involves continuously sharing critical information about one’s status.
M2 mentioned, ``\textit{The clear communication means where you are working, what you are doing, what your hazards are, and also letting other people know where you are. So I think poor communication is a standard causal factor for most accidents}''.
However, miscommunication is common with workers sometimes completing training sessions without fully understanding protocols (M1).
Lack of proper instructions and leadership is also related to this challenge, as W1 mentioned, ``\textit{The construction industry is really bad at how they promote people, they don’t ever give them the training to be a good leader}''.

\subsubsection{\textit{Low Workplace Empathy}}
Social bonding presents another challenge for individuals with ADHD, as R2 said, ``\textit{They may fear an increased risk of social rejection if they perceive their coworkers as unsupportive or lacking empathy}''.
W2 faced a dilemma, ``\textit{I explained my ADHD to one of my team mates, he gets it, but I tried to explain it to the lead person above him, he’s very hardheaded, stubborn, and doesn’t care}''.
A lack of understanding from neurotypical coworkers, coupled with the absence of acknowledgment or accommodation for ADHD-related challenges, can deepen feelings of isolation, as W1 mentioned, ``\textit{Some days I feel invisible. People seem indifferent to my struggle. I can’t quit without another job lined up because I need the money}''.

\subsection{\textbf{Productivity Support Strategies}}

\subsubsection{\textit{Clear Task Guidance}}
Explicit and step-by-step instructions are important for ADHD workers who struggle with attention to detail.
W1 mentioned, ``\textit{If workers can get a thorough understanding of the task, such as ‘Where does this pipe start? Where does it end? What does the pipe venting' that can be helpful}''.
Participants also emphasized that early exposure and preparation are essential to making workers more comfortable on construction sites. 
Thorough early preparation and upfront task guidance were described as foundational to project success and for reducing task ambiguity (M2).
Observing others at work provides ADHD workers with concrete models of task execution.
As W2 mentions, \textit{``Watching my teammate doing his job, I think of him as my model and it helps me going..I always needed a different way to learn than other people}''.
This modeling allows them to better understand sequencing, timing, and physical coordination, making it easier to replicate tasks accurately.
Participants emphasized the importance of structured guidance to help workers with ADHD understand task goals and maintain focus throughout task execution.
For example, M1 noted, \textit{``The more that you could expose folks in advance, make them feel more comfortable once they got out in a construction project''}. 
Others suggested (W2, R2) incorporating narration or contextual explanations to support comprehension during task execution.
\subsubsection{\textit{Continuous Monitoring Support}}
Participants highlighted the importance of instant feedback and periodic check-ins in improving task performance (W1, R1).
Without timely feedback, individuals with ADHD may drift off task or become overwhelmed by uncertainty.
Monitoring progress, following up on mistakes, and feedback during safety training, particularly in high-risk environments i.e., construction, is useful.
M2 commented on a VR video clip shown during the interview, ``\textit{The feature that does for fall protection, where you have to pick the right length of lanyard then it tells you did it right. When you did it wrong, it shows the animation hitting the ground. So it gives that instant feedback that helps}''.
From VR design perspectives, participants consistently emphasized the need for systems that can actively capture and redirect attention when needed. 
Participants described scenarios where real-time prompts could prevent errors or unsafe behaviors, such as reminders to check equipment or follow safety procedures. 
As R1 stated, \textit{``Hey, did you check this? You need to check and make sure that the guards are on''}.
\subsubsection{\textit{Collaborative Scaffolding}}
Companionship provides ADHD workers with an external source of accountability and motivation that can help maintain focus during demanding tasks~\cite{barkley2015executive}.
R1 explained how collaborative environments provide built-in support systems that can enhance the performance of workers with ADHD - ``\textit{working collaboratively can also be kind of fueling for them. Collaborate environments often have a support structure. I can model off someone else which helps me know what to do. Someone else can tap me on the shoulder when I'm zoned out}''.
Peer-guided hands-on learning is especially helpful for workers with ADHD, rather than relying solely on lengthy verbal explanations.
For instance W2 remarked, ``\textit{He also has ADHD and he won't explain things much, but he'll kind of guide me through things, and do the task a bunch of times in front of me. And if I have any questions. He answers them}''.
Participants also identified the need for mechanisms that encourage workers to periodically pause and evaluate their progress. 
Such reflective practices can help individuals with ADHD stay aligned with task requirements and reduce errors caused by inattention. 

\subsubsection{\textit{Repetitive Task Advantage}}
Although ADHD is commonly associated with distractibility, individuals with ADHD can excel in structured and repetitive tasks that provide clear expectations (R2). 
Repetitive tasks reduce the need for constant decision-making or shifting attention and minimize executive functioning demands~\cite{barkley2015executive}.
For instance, W1 mentioned, ``\textit{I guess a repetitive job is more helpful for me than multitasking, I'm repeating it to myself as I'm doing it}''.
M1 also said, ``\textit{I think that construction can be a very good environment for someone with ADHD, because some tests are repetitive in nature and allow them to function as opposed to needing to put multiple things together at the same time}''.
Providing a stable and predictive structure, these types of tasks reduce the cognitive burden of task switching and complex decision-making.

\subsubsection{\textit{Gamified Motivation}}
Creating an environment that incorporates fun and lighthearted interaction can significantly improve engagement for workers with ADHD.
R2 mentions, ``\textit{I think, having fun on the job is very important, like talking casual, bantering back and forth, and just having a good relationship with the people that you work with}''.
Participants highlighted the importance of tracking performance and providing motivational incentives to sustain engagement. 
Gamified elements, such as scoring systems and rankings, were seen as effective ways to encourage focus and continuous improvement. 
Such mechanisms can help users monitor their progress and set performance goals.
W2 also pointed how healthy competition can act as a motivator in the workplace, ``\textit{I think, having competing or like somewhat of a competition, is healthy. how fast you did it, and how few mistakes you got.. You'll get like an A}''.
Instead of viewing competition as pressure, it can be considered as a way to increase productivity and encourage focus.

%% file: sections/05_design.tex
\section{Discussion}

Our findings highlight a fundamental mismatch between the cognitive demands of construction work and the attentional and self-regulatory needs of workers with ADHD.
Construction sites are inherently dynamic, requiring workers to adapt to frequent procedural changes, and for workers with ADHD, these demands amplify their intrinsic challenges.
Hence, it is important to design assistive support systems that are sensitive to the nature of tasks performed in construction environments.
In this context, AI can take on multiple roles: as a decision-support system that assists workers under time pressure, as an environmental cueing mechanism that enhances situational awareness, and as a self-supportive or agentic assistant that provides reminders and scaffolding.
Our findings further suggest that combining context-aware sensing (e.g., gaze, task progress) with user-driven inputs (e.g., self-report or interaction cues) can enable more effective and adaptive support, aligning with prior work on mixed-initiative systems that balance automation with human control~\cite{allen1999mixed, horvitz1999principles}.

Beyond AI, our results also contribute to the design of VR for construction contexts. 
Existing literature primarily positions VR as a tool for safety training and skill development, demonstrating its effectiveness in improving hazard recognition and situational awareness in controlled environments~\cite{wang2018critical, li2018critical}. 
Our findings extend this perspective by identifying VR as a potential medium for attentional support, particularly for workers with ADHD, an area that remains largely unexplored. 
Participants highlighted how VR can provide structured guidance, interactive feedback, and immersive simulations that reduce cognitive load and enhance engagement. 
However, it is important to distinguish between training-time applications and field deployment. 
Immersive systems in live construction settings raise safety concerns, such as reduced peripheral awareness, occlusion of hazards, and increased sensory load~\cite{kourtesis2020virtual}.
These risks may be especially pronounced for individuals with ADHD, who are more sensitive to distraction and overstimulation. 
Therefore, VR-based interventions should be carefully scoped. 
Instead of fully immersive systems during live tasks, more lightweight, peripheral, or non-occluding assistive technologies may be safer and more appropriate for real-world construction contexts.

AI-enabled VR systems can serve as tools for building both individual resilience and adaptive capacity by providing scaffolding that helps workers navigate complex tasks, recover from errors, and maintain performance under pressure.
However, our results indicate that adaptability is not solely an individual responsibility but is shaped by organizational and social contexts.
Challenges such as coordination gaps, low workplace empathy, and inadequate communication structures suggest that systemic factors play a significant role in shaping workers’ experiences.
AI-enabled VR systems, therefore, should not be limited to individual assistance but should also incorporate features that enhance communication, coordination, and team-level situational awareness~\cite{endsley2017situation}.

\subsection{Design Implications}

Through our study, we identified key challenges faced by workers with ADHD in their workplaces, as well as forms of support that could enhance their performance. 
By synthesizing insights from the study, we derived design considerations that informed the development of appropriate evaluation tasks and attentional support interventions to explore.
Table 2 shows the mapping between the findings and design considerations.

\begin{table}[t]
\label{tab:design_implications}
\begin{tabularx}{\textwidth}{p{0.10\textwidth} p{0.32\textwidth} X}
\toprule
\textbf{Theme} & \textbf{Finding} & \textbf{Design Implication} \\
\midrule
\multirow{4}{=}{\small{Workplace Challenges}} 
& \small{Decision-making under pressure} 
& \small{Real-time AI decision support (~\ref{DI1} )} \\

& \small{Attention \& cognitive burden} 
& \small{Step-by-step guidance, reduce task complexity (~\ref{DI2} )} \\

& \small{Safety issues/missed hazards} 
& \small{Add context-aware visual/audio alerts (~\ref{DI1} )} \\

& \small{Procedural adaptivity} 
& \small{Add task update alerts, provide visual cues (~\ref{DI1},~\ref{DI2} ) } \\

& \small{Coordination gaps}
& \small{Shared task awareness and clear instructions (~\ref{DI2},~\ref{DI3} )} \\

\midrule
\multirow{5}{=}{\small{Productivity Support Strategies}} 
& \small{Need for clear guidance}
& \small{Visual structured workflows, voice narration (~\ref{DI2} )}\\

& \small{Continuous monitoring support}
& \small{Provide instant, adaptive feedback (~\ref{DI1}, ~\ref{DI3} ) } \\

& \small{Collaborative support}
& \small{Enable avatars/AI agents for co-working presence (~\ref{DI3} )} \\

& \small{Repetition enabled focus}
& \small{Design predictable task with repetitive steps (~\ref{DI2} ) } \\

& \small{Motivation through fun/competition}
& \small{Include gamification scores, rewards (~\ref{DI4} )} \\
\bottomrule
\end{tabularx}

\caption{
Mapping between Theme findings and Design Implications
}
\end{table}

\subsubsection{Context-Aware Attentional Interventions for Safety-Critical Moments}
\label{DI1}
For individuals with ADHD, divided attention presents a significant challenge, particularly in high-risk settings (i.e., construction sites) where situational awareness requires monitoring both primary tasks and surrounding hazards.
The ability to shift attention and re-engage with the primary task is essential for successful multitasking and situational awareness.
In VR, introducing moving objects around the player can effectively simulate unpredictable on-site events, such as passing equipment or falling materials.
AI systems could leverage some behavioral user data, such as head and gaze direction, task completion time, error frequency, idle duration, and response latency to hazards. 
When a critical situation appears in the VR simulation, the system could detect indicators of attentional drift (e.g., sustained gaze away from task-relevant areas), cognitive overload, such as clustered errors, or disengagement (e.g., prolonged inactivity). 
Based on these assessments, the AI could make constrained, context-aware adjustments—such as enhancing hazard salience, triggering visual prompts, amplifying motion cues, delivering spatialized audio effects, modulating task pacing, or temporarily reducing concurrent stimuli.

\subsubsection{Leveraging Repetition with AI-Guided Task Structuring}
\label{DI2}
We found that individuals with ADHD can perform well on repetitive tasks in environments with clear routines, as repetition reduces the burden of constant decision-making. 
However, it is essential that repetition is not purely monotonous but instead supported by external visual hints to help ADHD workers remain attentive and minimize errors.
Visual cues act as external scaffolds that reduce the need for constant self-monitoring and provide immediate guidance~\cite{shayesteh2022enhanced,hu2024exploring,cuber2024examining}.
For instance, AI can provide a visual reference model or pattern that allows participants to build something through repetitive actions, while still demanding attentional engagement. 
AI can also provide structured, step-by-step task decomposition guidelines.
During sudden procedural changes, AI-mediated VR systems can provide stepwise visual progression cues that dynamically guide users through evolving task sequences.
Instead of presenting a static list of instructions, systems can visually highlight the current step while automatically marking completed steps (e.g., crossing them off or fading them out). 
This approach not only clarifies what has been done and what remains but also offers an intrinsic sense of progress, which can be motivating and reduce cognitive overload.
Such designs blend predictability with attentional challenge, creating a balance that supports both focus and performance.

\subsubsection{Providing Sense of Accountability and Shared Progress}
\label{DI3}
Participants noted that watching others perform a task helps them both learn and stay engaged in their own work. 
When workers can track not only their own progress but also that of others, it creates a comparative frame of reference that anchors attention and sustains motivation~\cite{kuntsi2011intraindividual}. 
In VR, this can be facilitated by placing players in view of one another where they can see each other, or by incorporating progress bars that display task status. 
These features function as a reminder for continued focus and effort.
Co-working arrangements, such as body doubling~\cite{annavarapu2024comparative}, establish an immediate sense of accountability through the presence of another person, helping individuals with ADHD remain anchored to their tasks. 
In VR, this presence can be represented through avatars, characters, or even ambient supportive AI agents such as NPCs (Non-Playable Characters), which can provide a sense of shared engagement.
The agent can facilitate real-time companionship, provide real-time corrective feedback to users during task execution, as well as post-task performance summaries that highlight strengths and areas for improvement.

\subsubsection{Introduce Motivational Dynamics for Engagement}
\label{DI4}
We found that making the workplace enjoyable and motivating is essential for enhancing productivity.
AI-mediated VR systems can facilitate mechanisms, i.e., visible rankings, progress status, or reward systems for achieving goals, which can sustain engagement and encourage persistence. 
For collaborative tasks, the system can incorporate performance tracking, and feedback loops based on speed, accuracy, and safety compliance.
However, such motivational mechanisms must be carefully designed to account for individual differences among workers with ADHD.
Although comparative elements such as visible rankings or scoreboards may motivate some individuals, they may also introduce pressure, social comparison, or feelings of isolation for others, particularly in environments where neurodiversity is not well understood. 
In those situations, systems should prioritize personalized and self-referenced feedback, allowing workers to track their own progress over time rather than comparing performance with others.
Additionally, incorporating adaptive motivational cues, such as individualized encouraging quotes, goal-setting, and non-competitive rewards, can provide immediate reinforcement for sustained engagement.

%% file: sections/06_conclusion.tex
\section{Limitations}

This study has several limitations. 
The small sample size limits the generalizability of the findings, particularly given that ADHD is a spectrum condition and the challenges of recruiting construction workers with clinically diagnosed ADHD in real-world settings. 
Although the study achieved thematic sufficiency in identifying cross-cutting patterns, the inclusion of diverse stakeholders to capture multiple perspectives means that our findings should be interpreted as exploratory rather than representative.
As such, this work is best positioned as a preliminary or pilot study that surfaces early insights into ADHD-related challenges and opportunities for AI- and VR-based interventions. 
Future research with larger samples and in-situ or longitudinal methods is needed to validate and extend these findings.

\section{Conclusions and Future Work}

In this work, we explore how workers with ADHD experience and manage the demands of construction work through an interview study with ADHD workers, safety managers, and ADHD researchers. 
Our findings identify key challenges, including decision making, procedural adaptivity, and safety culture stagnation for construction workers with ADHD.
We also highlight useful strategies like structured visual guidance, shared monitoring, and repetition that can support focus and productivity.
We further demonstrate the potential of AI-enabled VR systems to provide adaptive, context-aware support through real-time guidance and feedback.
In the future, we plan to extend this work to more deeply investigate ADHD-related workplace challenges across other safety and mission-critical environments and explore design opportunities for next-generation XR systems.